\documentclass[12pt]{revtex4-2}  

\usepackage{amsmath}  
\usepackage{amsfonts} 
\usepackage{graphicx} 
\usepackage{xcolor}
\usepackage{mathrsfs}
\usepackage{array}
\usepackage{makecell}

\begin{document}


\title{Mechanical Attributes of Fractal Dragons
}


\author{Huy T. Q. Phan}
\thanks{These authors contributed equally to this work.}
\affiliation{University of Technology - VNUHCM,\\268 Ly Thuong Kiet, Ho Chi Minh 700000, Vietnam.}

\author{Duc M. Bui}
\thanks{These authors contributed equally to this work.}
\affiliation{École Polytechnique, Route de Saclay, 91128 Palaiseau, France.}

\author{Cong T. Than}
\affiliation{Bac Giang Specialized Upper Secondary School, \\ 76J4+QvQ, Dinh Ke, Bac Giang 24000, Vietnam.}

\author{Trung V. Phan}
\email{tphan23@jhu.edu}
\affiliation{Department of Chemical and Biomolecular Engineering, \\ John Hopkins University, Baltimore, MD 21218, USA.}

\begin{abstract}
Fractals are ubiquitous natural emergences that have gained increased attention in engineering applications, thanks to recent technological advancements enabling the fabrication of structures spanning across many spatial scales. We show how the geometries of fractals can be exploited to determine their important mechanical properties, such as the first and second moments, which physically correspond to the center of mass and the moment of inertia, using a family of complex fractals known as the dragons.
\end{abstract}

\date{\today}

\maketitle 

\section{Fractals and Dragons}

Beno\^it B. Mandelbrot has elegantly described how fractals can be seen everywhere in nature with these words, ``clouds are not spheres, mountains are not cones, coastlines are not circles, and bark is not smooth, nor does lightning travel in a straight line'' \cite{mandelbrot1982fractal}. These pervasive patterns, deviating from traditional geometric norms \cite{barnsley2014fractals} and displaying self-similarity across various scales \cite{falconer2004fractal}, often emerge through the chaotic or stochastic self-organization of underlying dynamics  \cite{rigon1994landscape,rodriguez1997fractal,isaeva2012self}.  From an engineering perspective, despite their apparent complexity, fractals are easy to create with high-resolution fabrication technology \cite{phan2020bacterial}, as the same patterns are repeated but at different length scales. A fractal machine can operate on the same underlying physical principles across all spatial scales, making calibration and maintenance remarkably straightforward \cite{li2023inertial}. Mother Nature has used fractal geometries for hundreds of millions of years to optimize biological functions \cite{kurakin2011self}, and only in recent times have humans started emulating these for innovative device designs and developments \cite{brambila2017fractal}.

There have been numerous applications of utilizing fractals in mechanics \cite{fan2014fractal}, thermodynamics \cite{ochoa2015optimization}, hydrodynamics \cite{xie2020experimental}, electronics \cite{fairbanks2011fractal}, electromagnetics \cite{werner2003overview}, and more \cite{segev2012fractal}. Distinct topologies give rise to distinct physical manifestations \cite{phan2021curious,ao2023schrodinger}, and phenomena emerging within fractal spaces not only deviate significantly but also introduce new behaviors compared to their expected ones in conventional Euclidean spaces. Thus, to engineer fractals, it is crucial to understand the unconventional physical properties that can arise from these \cite{phan2024vanishing}. Here, we investigate the mechanical properties of fractals, specifically focusing on fundamental attributes i.e. the first and second moments. 
These quantities, which correspond physically to the center of mass and the moments of inertia, represent the basic mechanical characteristics of geometric shapes or even biological entities \cite{santschi1963moments}. We consider a family of two-dimensional space fractals called dragons, chosen for their inherent complexity, which allows us to showcase the general applicability of our approach in calculating the moments of these infinite-intricacy structures. 

\begin{figure*}[!htbp]
\includegraphics[width=\textwidth]{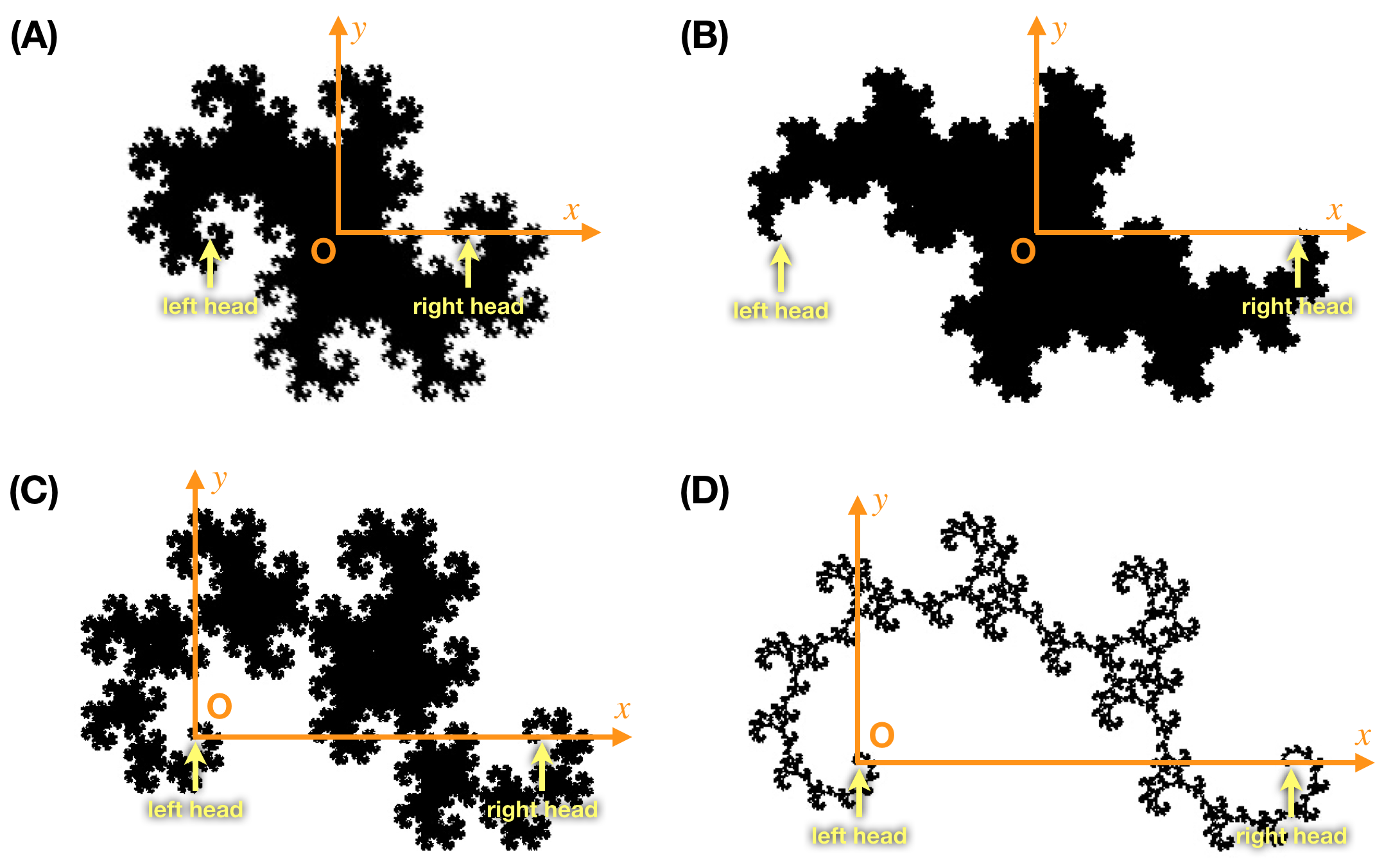}
\caption{\textbf{Fractal dragons.} Each dragon has two heads at normalized distance $L\equiv 1$ apart, and we choose a coordinates system O$xy$ so that O$x$ passes through both heads. \textbf{(A)} Twindragon. \textbf{(B)} Terdragon. \textbf{(C)} Heighway dragon. \textbf{(D)} Golden dragon. The origin O is the middle of two heads for every symmetric dragon \textbf{(A,B)}, and is the left head for every assymmetric dragon \textbf{(C,D)}.}
\label{fig01}
\end{figure*}

{\color{black} Let us discuss the dragons we are going to deal with.} Dragon fractals are self-similar geometric shapes that can divide into instances of the very same pattern with orientations different from the original \cite{tabachnikov2014dragon}. In other words, a dragon can be recreated by assembling {\color{black} a finite number of} smaller dragons of various sizes through translational and rotational transformations. We show that, from these geometrical transformations and physical dimensional analysis, it is possible to obtain the exact analytical answers for the first and second moments of dragons without summing up an infinite series encoding how these dragons can be generated mathematically e.g. by a computer program. {\color{black} This is possible because these moments transform as tensors under rotation, and these fractals possess well-defined self-similar dimensionalities under rescaling.} We demonstrate our method with two examples, the symmetric twindragon \cite{agnesscottTwindragon} and the asymmetric golden dragon \cite{agnesscottGoldenDragon}. We also give answers to other interesting dragon fractals, i.e. the tedragon \cite{agnesscottTerdragon} and the Heighway dragon \cite{agnesscottHeighwayDragon}. We then verify our findings through simple estimations using direct numerical calculations for these four dragon fractals (see Fig. \ref{fig01}).

\ \ 

\section{Center of Mass and Moments of Inertia}

The center of mass and the moments of inertia offer simplified mechanical descriptions of a rigid body. In many physical settings, the translational motion of the rigid body, influenced by external forces, can be accurately approximated as if it were concentrated at a single point, which is the center of mass. The response to rotational motion, induced by external torques, can also be reliably predicted by knowing the moments of inertia. 

\subsection{Mathematical Definitions \label{CoM_MoI_defs}}

Consider a two-dimensional rigid body $\Omega$ which is a collection of constituents each labelled $j$, each associated with weight $w_j$ and located at position $[x_j,y_j]$ in a Cartesian coordinate system O$xy$. Let $W(\Omega) \equiv \sum_j w_j$ be the total weight of this rigid body. The center of mass is defined to be the point G $[x_\text{G},y_\text{G}]$ which can be calculated from the first-order monomial of constituents position coordinates:
\begin{equation}
[x_\text{G},y_\text{G}] = \left[ X(\Omega), Y(\Omega) \right] \equiv \left[ \frac{\sum_j w_j x_j}{  \sum_j w_j} , \frac{  \sum_j w_j y_j}{  \sum_j w_j} \right] \ ,
\label{CoM_eq}
\end{equation}
When we split the rigid body $\Omega$ into two smaller ones, $\Omega_1$ and $\Omega_2$, each composed of constituents with weights $\{ w_{j_1} \}$ and $\{ w_{j_2} \}$ respectively, located at $\{ [x_{j_1}, y_{j_1}] \}$ and $\{ [x_{j_2}, y_{j_2}] \}$, i.e. $\Omega = \Omega_1 \cup \Omega_2$, an additive rule follows from this mathematical definition:
\begin{equation}
\left[ X(\Omega), Y(\Omega) \right] = \left[ \frac{ W(\Omega_1) X(\Omega_1) + W(\Omega_2) X(\Omega_2)}{W(\Omega)}, \frac{ W(\Omega_1) Y(\Omega_1) + W(\Omega_2) Y(\Omega_2)}{W(\Omega)}\right] \ .
\label{additive_CoM}
\end{equation}
The center of mass G$_{1}$ of $\Omega_{1}$ is at positions $\left[ x_\text{G$_{1}$},y_\text{G$_{1}$} \right] = \left[ X(\Omega_{1}),Y(\Omega_{1})\right]$, and a similar identification goes for the same mechanical attribute of $\Omega_2$.

For a general definition for the moments of inertia, we need to define it with respect to a point A $\left[ x_\text{A}, y_\text{A} \right]$. It is a matrix $\mathbb{I}(\Omega,\text{A})$ with three independent components:
\begin{equation}
\mathbb{I}(\Omega,\text{A}) = \begin{bmatrix}
I_{xx} (\Omega,\text{A}) & I_{xy} (\Omega,\text{A}) \\
I_{yx} (\Omega,\text{A}) & I_{yy} (\Omega,\text{A}) 
\end{bmatrix} \ , \ I_{xy} (\Omega,\text{A}) = I_{yx} (\Omega,\text{A})  \ .
\end{equation}
The formula for these are given by the second-order polynomials of constituents position coordinates:
\begin{equation}
\begin{split}
I_{xx} (\Omega,\text{A}) \equiv \sum w_j \left( y_j - y_\text{A} \right)^2  \ , \ I_{yy} (\Omega,\text{A}) \equiv \sum w_j \left( x_j - x_\text{A} \right)^2 \ , 
\\
I_{xy} (\Omega,\text{A}) \equiv -\sum w_j \left( x_j - x_\text{A} \right)\left( y_j - y_\text{A} \right) \ .
\end{split}
\label{MoI_eq}
\end{equation}
The additive rule for general moments of inertia is quite straight forward:
\begin{equation}
\mathbb{I}(\Omega,\text{A}) = \mathbb{I}(\Omega_1,\text{A}) + \mathbb{I}(\Omega_2,\text{A}) \ .  
\end{equation}
In this work, whenever we refer to the moments of inertia without specifying any point on space, we means it is with respect to the center of mass G. We use the following notations:
\begin{equation}
\tilde{\mathbb{I}}(\Omega) = \begin{bmatrix}
\tilde{I}_{xx} (\Omega) & \tilde{I}_{xy} (\Omega) \\
\tilde{I}_{yx} (\Omega) & \tilde{I}_{yy} (\Omega) 
\end{bmatrix} \equiv \mathbb{I}(\Omega,\text{G}) \ .
\end{equation}
From this definition, we obtain the parallel axis theorem \cite{morin2008introduction}.
\begin{equation}
\begin{split}
I_{xx} (\Omega,\text{A}) = \tilde{I}_{xx} (\Omega) + W(\Omega) \left( y_\text{A}-y_\text{G}\right)^2  \ , \ I_{yy} (\Omega,\text{A}) = \tilde{I}_{yy} (\Omega) + W(\Omega) \left( x_\text{A}-x_\text{G}\right)^2  \ , 
\\
I_{xy} (\Omega,\text{A}) = \tilde{I}_{xx} (\Omega) - W(\Omega) \left( x_\text{A}-x_\text{G}\right) \left( y_\text{A}-y_\text{G}\right) \ .
\end{split}
\end{equation}
Combining this with the additive property $\tilde{\mathbb{I}}(\Omega) = \mathbb{I}(\Omega_1,\text{G}) + \mathbb{I}(\Omega_2,\text{G})$ gives:
\begin{equation}
\begin{split}
\tilde{I}_{xx} (\Omega) &= \tilde{I}_{xx} (\Omega_1) + \tilde{I}_{xx} (\Omega_2) + W(\Omega_1) \left( y_\text{G$_1$}-y_\text{G}\right)^2 + W(\Omega_2) \left( y_\text{G$_2$}-y_\text{G}\right)^2  \ ,
\\
\tilde{I}_{yy} (\Omega) &= \tilde{I}_{yy} (\Omega_1) + \tilde{I}_{yy} (\Omega_2) + W(\Omega_1) \left( x_\text{G$_1$}-x_\text{G}\right)^2 + W(\Omega_2) \left( x_\text{G$_2$}-x_\text{G}\right)^2  \ ,
\\
\tilde{I}_{xy} (\Omega) &= \tilde{I}_{xy} (\Omega_1) + \tilde{I}_{xy} (\Omega_2)
\\
& \ - W(\Omega_1) \left( x_\text{G$_1$}-x_\text{G}\right) \left( y_\text{G$_1$}-y_\text{G}\right) - W(\Omega_2) \left( x_\text{G$_2$}-x_\text{G}\right) \left( y_\text{G$_2$}-y_\text{G}\right) \ .
\end{split}
\label{useful_rel}
\end{equation}
These are very useful relations for calculating the moments of inertia of fractal dragons.

If all weights $\left\{ w_j \right\}$ are equal and the total weight is normalized to $W \equiv 1$, then the center of mass and the moments of inertia become the first and second moments:
\begin{equation}
\left[X(\Omega),Y(\Omega)\right] = \left[ \langle x\rangle , \langle y \rangle \right] \ , \ \tilde{\mathbb{I}}(\Omega) = \begin{bmatrix}
\left\langle \left( y - \langle y \rangle \right)^2 \right\rangle & -\left\langle \left( x - \langle x \rangle \right)\left( y - \langle y \rangle \right) \right\rangle \\
-\left\langle \left( x - \langle x \rangle \right)\left( y - \langle y \rangle \right) \right\rangle & \left\langle \left( x - \langle x \rangle \right)^2 \right\rangle 
\end{bmatrix} \ .
\label{1st_and_2nd}
\end{equation}
in which $\langle \bullet \rangle$ denotes averaging over the set of values $\left\{\bullet_j \right\}$. We use these simplifications for the rest of this work.

\subsection{Under Geometric Transformations}

The center of mass and the moments of inertia can change after geometric transformations apply on the object $\Omega$. Here we only consider scale transformation $\mathcal{S}_{\eta,\text{O}}$ a length factor $\eta$ with respect to the origin O {\color{black} (in which $\eta = 1$ when the spatial size remains unchanged, $\eta < 1$ when the spatial size after rescaling is smaller, and $\eta > 1$ when the spatial size after rescaling is larger)}, rotational transformation $\mathcal{R}_{\theta,\text{O}}$ an angle $\theta$ with respect to the origin O, and translational transformation $\mathcal{T}_{\left[ \lambda_x,\lambda_y\right]}$ a spatial-vector $\left[ \lambda_x,\lambda_y \right]$. These transformations act on spatial-coordinates $\left[ x,y\right]$ as follows:
\begin{equation}
\begin{split}
\mathcal{S}_{\eta,\text{O}} \left[ x,y\right] &= \left[ \eta x, \eta y\right] \ , 
\\
\mathcal{R}_{\theta,\text{O}} \left[ x,y\right] & = \left[ x\cos\theta + y\sin\theta , -x\sin\theta +y\cos\theta \right] \ ,
\\
\mathcal{T}_{\left[ \lambda_x,\lambda_y\right]} \left[ x,y\right] &= \left[ x + \lambda_x , y+\lambda_y \right] \ . 
\end{split}
\label{transf_coordinates}
\end{equation}
We define the way they act on the total weight of an object with similarity dimension $\mathscr{D}$:
\begin{equation}
W\left( \mathcal{S}_{\eta,\text{O}} \Omega \right) = \eta^{\mathscr{D}} \ , \ W\left( \mathcal{R}_{\theta,\text{O}} \Omega \right) = W \left( \mathcal{T}_{\left[ x',y'\right]} \Omega \right) = 1 \ ,    
\label{transf_weight}
\end{equation}
so that the ``weight density'' in the continuum limit is kept fixed. {\color{black} In simpler terms, in a high-resolution pixelated representation of the shape, each pixel containing the fractal curves is assigned an equal weight. This is consistent with the definition of the box-counting dimension \cite{yu2005coarse,schleicher2007hausdorff}, in which the similarity dimension $\mathscr{D}$ for dragon curves are calculated for. It should be noted that these values $\mathscr{D}$ may not be an integer, indicating fractal characteristics.}

To find the actions of these transformations on the center of mass and the moments of inertia, we just have to apply the transformations of spatial-coordinates and weights as described Eq. \eqref{transf_coordinates} and Eq. \eqref{transf_weight} on their mathematical definitions, which are given in Section \ref{CoM_MoI_defs}. Acting on the spatial-coordinates of center of mass gives identical results with those in Eq. \eqref{transf_coordinates}:
\begin{equation}
\begin{split}
 \left[ X(\mathcal{S}_{\eta,\text{O}} \Omega),Y(\mathcal{S}_{\eta,\text{O}} \Omega)\right] &= \left[ \eta X(\Omega), \eta Y(\Omega)\right] \ , 
\\
 \left[ X(\mathcal{R}_{\theta,\text{O}}\Omega),Y(\mathcal{R}_{\theta,\text{O}} \Omega)\right]  & = \left[ X(\Omega)\cos\theta + Y(\Omega) \sin\theta , -X(\Omega) \sin\theta + Y(\Omega)\cos\theta \right] \ ,
\\
 \left[ X(\mathcal{T}_{\left[ \lambda_x,\lambda_y\right]} \Omega),Y(\mathcal{T}_{\left[ \lambda_x,\lambda_y\right]} \Omega)\right] &= \left[ X(\Omega) + \lambda_x , X(\Omega) +\lambda_y \right] \ . 
\end{split}
\label{transf_CoM}
\end{equation}
Now, let us focus at each of these transformations separately when acting on the moments of inertial, we find the followings:
\begin{itemize}
    \item \textbf{Scale transformation:} There is an overall factor after the change, coming from both the transformations of weights (linear) and spatial-coordinates (quadratic):
    \begin{equation}
    \tilde{\mathbb{I}} (\mathcal{S}_{\eta,\text{O}} \Omega) = \eta^{\mathscr{D}+2} \tilde{\mathbb{I}} ( \Omega) \ .
    \label{transf_scale_MoI}
    \end{equation}
    {\color{black} We explain this factor in Appendix \ref{scale_transf}.}
    \item \textbf{Rotational transformation:} The matrix $\tilde{\mathbb{I}}$ transforms like a tensor \cite{morin2008introduction}:
    \begin{equation}
    \begin{split}
    \tilde{I}_{xx}\left(\mathcal{R}_{\theta,\text{O}} \Omega \right) &= \tilde{I}_{xx}(\Omega) \cos^2\theta +\tilde{I}_{yy}(\Omega) \sin^2\theta + 2\tilde{I}_{xy}(\Omega) \cos\theta \sin\theta \ ,
    \\
    \tilde{I}_{yy}\left(\mathcal{R}_{\theta,\text{O}} \Omega \right) &= \tilde{I}_{xx} (\Omega) \sin^2\theta +\tilde{I}_{yy} (\Omega) \cos^2\theta - 2\tilde{I}_{xy} (\Omega) \cos\theta \sin\theta \ ,
    \\
    \tilde{I}_{xy}\left(\mathcal{R}_{\theta,\text{O}} \Omega \right) &= -\tilde{I}_{xx} (\Omega) \sin\theta \cos\theta + \tilde{I}_{yy} (\Omega) \sin\theta \cos\theta +\tilde{I}_{xy} (\Omega) \left( \cos^2\theta - \sin^2\theta \right) \ .
    \label{transf_rotate_MoI}
    \end{split}
    \end{equation}
    \item \textbf{Translational transformation:} No change emerges:
    \begin{equation}
    \tilde{\mathbb{I}} (\mathcal{T}_{\left[ \lambda_x,\lambda_y\right]} \Omega) = \tilde{\mathbb{I}} ( \mathcal{T}_{\left[ \lambda_x,\lambda_y\right]} \Omega) \ .
    \label{transf_translate_MoI}
    \end{equation}
\end{itemize}
We now have all the basic ingredients to find the first and second moments of fractal dragons.

Define a sequence of transformations, proceeding in the order of scaling, rotation, and translation:
\begin{equation}
\mathcal{C}_{\left[ \lambda_x,\lambda_y\right];\theta;\eta} \equiv \mathcal{T}_{\left[ \lambda_x,\lambda_y\right]} \mathcal{R}_{\theta,\text{O}} \mathcal{S}_{\eta,\text{O}} \ .
\label{transf_seq}
\end{equation}
For the total weight, we use Eq. \eqref{transf_weight}:
\begin{equation}
W\left( \mathcal{C}_{\left[ \lambda_x,\lambda_y\right];\theta;\eta} \Omega \right) = \eta^{\mathscr{D}} W\left( \Omega \right) \ .   
\label{transf_chain_weight}
\end{equation}
For the center of mass, we use Eq. \eqref{transf_CoM}:
\begin{equation}
\begin{split}
X\left( \mathcal{C}_{\left[ \lambda_x,\lambda_y\right];\theta;\eta} \Omega \right) &=  \eta X(\Omega)\cos\theta + \eta Y(\Omega) \sin\theta + \lambda_x \ ,
\\
Y\left( \mathcal{C}_{\left[ \lambda_x,\lambda_y\right];\theta;\eta} \Omega \right) &= - \eta X(\Omega) \sin\theta + \eta Y(\Omega)\cos\theta + \lambda_y \ .
\end{split}
\label{transf_chain_CoM}
\end{equation}
For the moments of inertial, we use Eq. \eqref{transf_scale_MoI}, Eq. \eqref{transf_rotate_MoI}, and Eq. \eqref{transf_translate_MoI}:
\begin{equation}
    \begin{split}
    \tilde{I}_{xx}\left( \mathcal{C}_{\left[ \lambda_x,\lambda_y\right];\theta;\eta} \Omega \right) &= \tilde{I}_{xx}(\Omega) \eta^{\mathscr{D}+2} \cos^2\theta +\tilde{I}_{yy}(\Omega) \eta^{\mathscr{D}+2} \sin^2\theta + 2\tilde{I}_{xy}(\Omega) \eta^{\mathscr{D}+2} \cos\theta \sin\theta \ ,
    \\
    \tilde{I}_{yy}\left( \mathcal{C}_{\left[ \lambda_x,\lambda_y\right];\theta;\eta} \Omega \right) &= \tilde{I}_{xx}  (\Omega) \eta^{\mathscr{D}+2} \sin^2\theta +\tilde{I}_{yy} (\Omega) \eta^{\mathscr{D}+2} \cos^2\theta - 2\tilde{I}_{xy} (\Omega) \eta^{\mathscr{D}+2} \cos\theta \sin\theta \ ,
    \\
    \tilde{I}_{xy}\left( \mathcal{C}_{\left[ \lambda_x,\lambda_y\right];\theta;\eta} \Omega \right) &= -\tilde{I}_{xx} (\Omega) \eta^{\mathscr{D}+2} \sin\theta \cos\theta + \tilde{I}_{yy} (\Omega) \eta^{\mathscr{D}+2} \sin\theta \cos\theta
    \\
    & \ \ \ \ \ \ +\tilde{I}_{xy} (\Omega) \eta^{\mathscr{D}+2} \left( \cos^2\theta - \sin^2\theta \right) \ .
    \label{transf_chain_MoI}
    \end{split}
    \end{equation}
For the rest of this work, we formulate all necessary transformations in this sequential format.

\section{Twindragon}

\begin{figure*}[!htbp]
\includegraphics[width=0.7\textwidth]{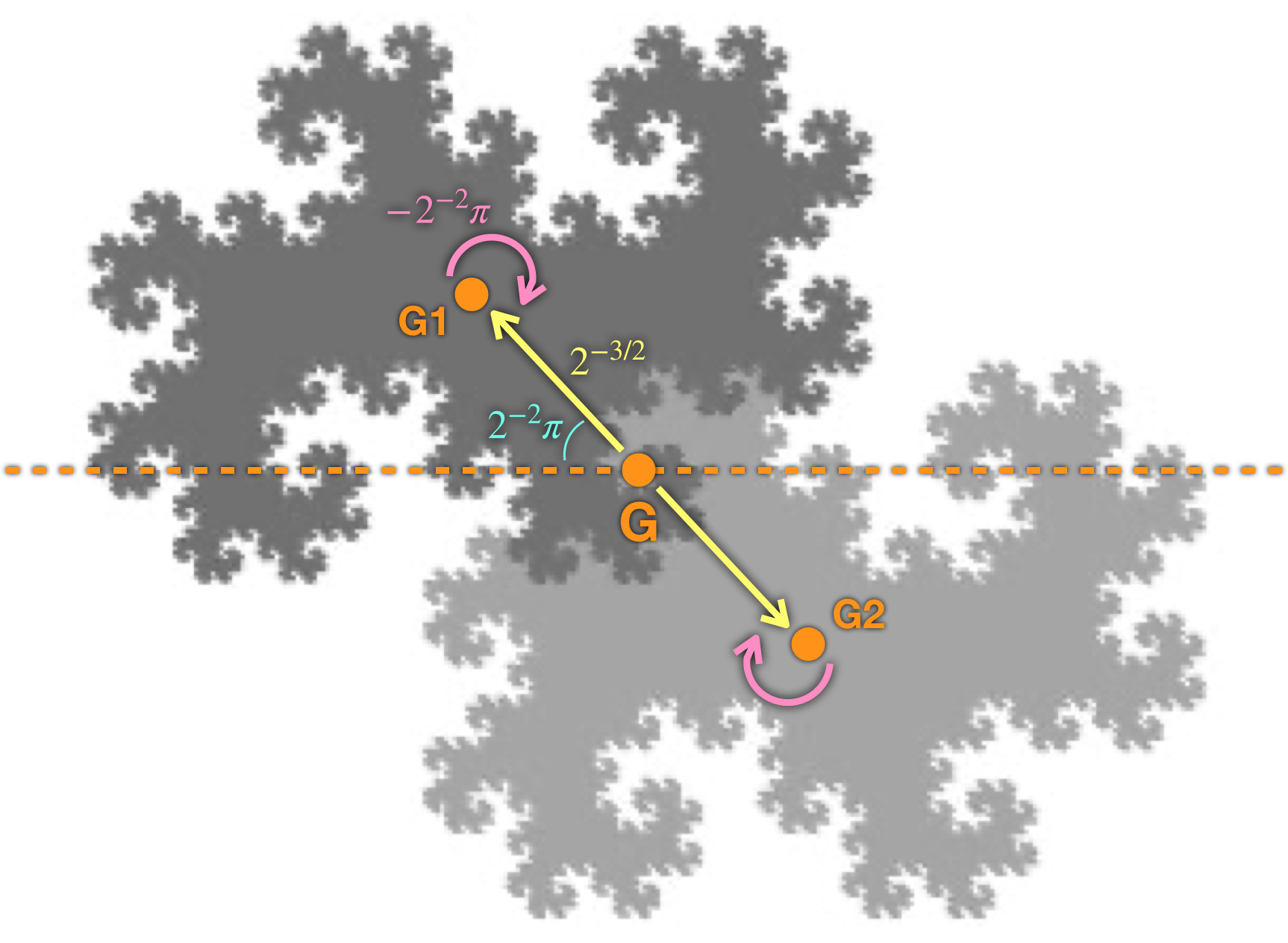}
\caption{\textbf{Twindragon.} A twindragon can be decomposed into two identical smaller twindragons as shown in two different shades of gray, which we call twindragon $1$ (on the left) and twindragon $2$ (on the right). Note that the rotation arrow indicates the chosen positive direction (clockwise), hence the angle values being negative here mean the smaller twindragons are rotated in the negative direction (counter-clockwise).}
\label{fig02}
\end{figure*}

Twindragon is a symmetric fractal of similarity dimension $\mathscr{D} = 2$, which indicates that it is a space-filling geometric structure. Choose a Cartesian coordinate system O$xy$ as in Fig. \ref{fig01}A, then due to symmetry the center of mass G is the same position as the origin O:
\begin{equation}
\left[ x_\text{G}, y_\text{G} \right] = \left[ 0,0 \right] \ .
\end{equation}
The distance between two heads of the dragon is normalized to $L\equiv 1$. A twindragon $\Omega^{(\text{Tw})}$ {\color{black} is self-similar, in which it} can be decomposed into two identical-size twindragons $\Omega^{(\text{Tw}_1)} \cup \Omega^{(\text{Tw}_2)}$, which we call twindragon $1$ and twindragon $2$ as in Fig. \ref{fig02}. From the geometrical information as given in \cite{agnesscottTwindragon}, we can determine that they have centers of mass located at positions $\left[ x_{\text{G}_1},y_{\text{G}_1}\right]=\left[ - 2^{-2}, 2^{-2}\right]$ and $\left[ x_{\text{G}_2},y_{\text{G}_2}\right]=\left[ 2^{-2}, - 2^{-2}\right]$ and are oriented at angles $\theta_1 = \theta_2 = -2^{-2}\pi$ away from the original twindragon. The distances between the two heads of these smaller dragons are $L_1=L_2=2^{-1/2}$. {\color{black} Both $\Omega^{(\text{Tw}_1)}$ and $\Omega^{(\text{Tw}_2)}$ has the same spatial rescaling factor i.e. $\eta_1 = L_1/L = 2^{-1/2}$ and $\eta_2 = L_2/L = 2^{-1/2}$, as compared to the original $\Omega^{(\text{Tw})}$. They also have equal weights, i.e. $ W\left(\Omega^{(\text{Tw}_1)}\right) = W\left(\Omega^{(\text{Tw}_2)}\right)= 1/2$ as follows from Eq. \eqref{transf_chain_weight} and the weight normalization $W\left(\Omega^{(\text{Tw})}\right)=1$.}

We can get the smaller twindragons from the original one by applying the following sequences of geometric transformation, as described in Eq. \eqref{transf_seq}:
\begin{equation}
\begin{split}
    \Omega^{(\text{Tw}_1)}  &= \mathcal{C}_{ [  x_{\text{G}_1},y_{\text{G}_1} ];\theta_1;L_1/L}  \Omega^{(\text{Tw})} = \mathcal{C}_{ [ -2^{-2},2^{-2}];-2^{-2}\pi;2^{-1/2}}  \Omega^{(\text{Tw})} \ ,
    \\
    \Omega^{(\text{Tw}_2)}  &= \mathcal{C}_{ [  x_{\text{G}_2},y_{\text{G}_2} ];\theta_2;L_2/L}  \Omega^{(\text{Tw})} = \mathcal{C}_{ [ 2^{-2},-2^{-2}];-2^{-2}\pi;2^{-1/2}}  \Omega^{(\text{Tw})} 
\end{split}
\label{Tw1_and_Tw2_transf}
\end{equation}
Take these to Eq. \eqref{transf_chain_weight} and Eq. \eqref{transf_chain_MoI}, then utilize the obtained results into Eq. \eqref{useful_rel} where we make the identifications $\Omega=\Omega^{(\text{Tw})}$, $\Omega_1=\Omega^{(\text{Tw}_1)}$, and $\Omega_2=\Omega^{(\text{Tw}_2)}$, we arrive at a set of three consistency equations with three unknowns:
\begin{equation}
\begin{split}
\tilde{I}_{xx} \left( \Omega^{(\text{Tw})} \right) &= 2^{-2} \tilde{I}_{xx} \left( \Omega^{(\text{Tw})} \right) + 2^{-2} \tilde{I}_{yy} \left( \Omega^{(\text{Tw})} \right) - 2^{-1} \tilde{I}_{xy} \left( \Omega^{(\text{Tw})} \right) + 2^{-4} \ ,
\\
\tilde{I}_{yy} \left( \Omega^{(\text{Tw})} \right) &= 2^{-2} \tilde{I}_{xx} \left( \Omega^{(\text{Tw})} \right) + 2^{-2} \tilde{I}_{yy} \left( \Omega^{(\text{Tw})} \right) + 2^{-1} \tilde{I}_{xy} \left( \Omega^{(\text{Tw})} \right) + 2^{-4} \ ,
\\
\tilde{I}_{xy} \left( \Omega^{(\text{Tw})} \right) &= 2^{-2} \tilde{I}_{xx} \left( \Omega^{(\text{Tw})} \right) - 2^{-2} \tilde{I}_{yy} \left( \Omega^{(\text{Tw})} \right) + 2^{-4} \ ,
\end{split}
\label{Tw_consistency_eqs}
\end{equation}
{\color{black} which can be derived as shown in Appendix \ref{Tw_consistency}. We then solve them} analytically:
\begin{equation}
\tilde{I}_{xx} = \frac1{10} = 0.1 \ , \ \tilde{I}_{yy} = \frac3{20} = 0.15 \ , \ \tilde{I}_{xy} = \frac1{20} = 0.05 \ .  
\end{equation}
These are the second moments of the twindragon fractal. 

We have employed a trick involving the formulation of consistency equations for physical systems of infinite complexity by decomposing them into repetitive components. This methodology, well-known among students of physics, has been exemplified and popularized through the familiar exercises of calculating the equivalent resistance of infinite resistor networks with e.g. the ladder topology \cite{leighton1965feynman} or even more intricated configurations \cite{nguyen2020infinite,tran2021elementary}.

\ \ 

\section{Golden Dragon}

\begin{figure*}[!htbp]
\includegraphics[width=\textwidth]{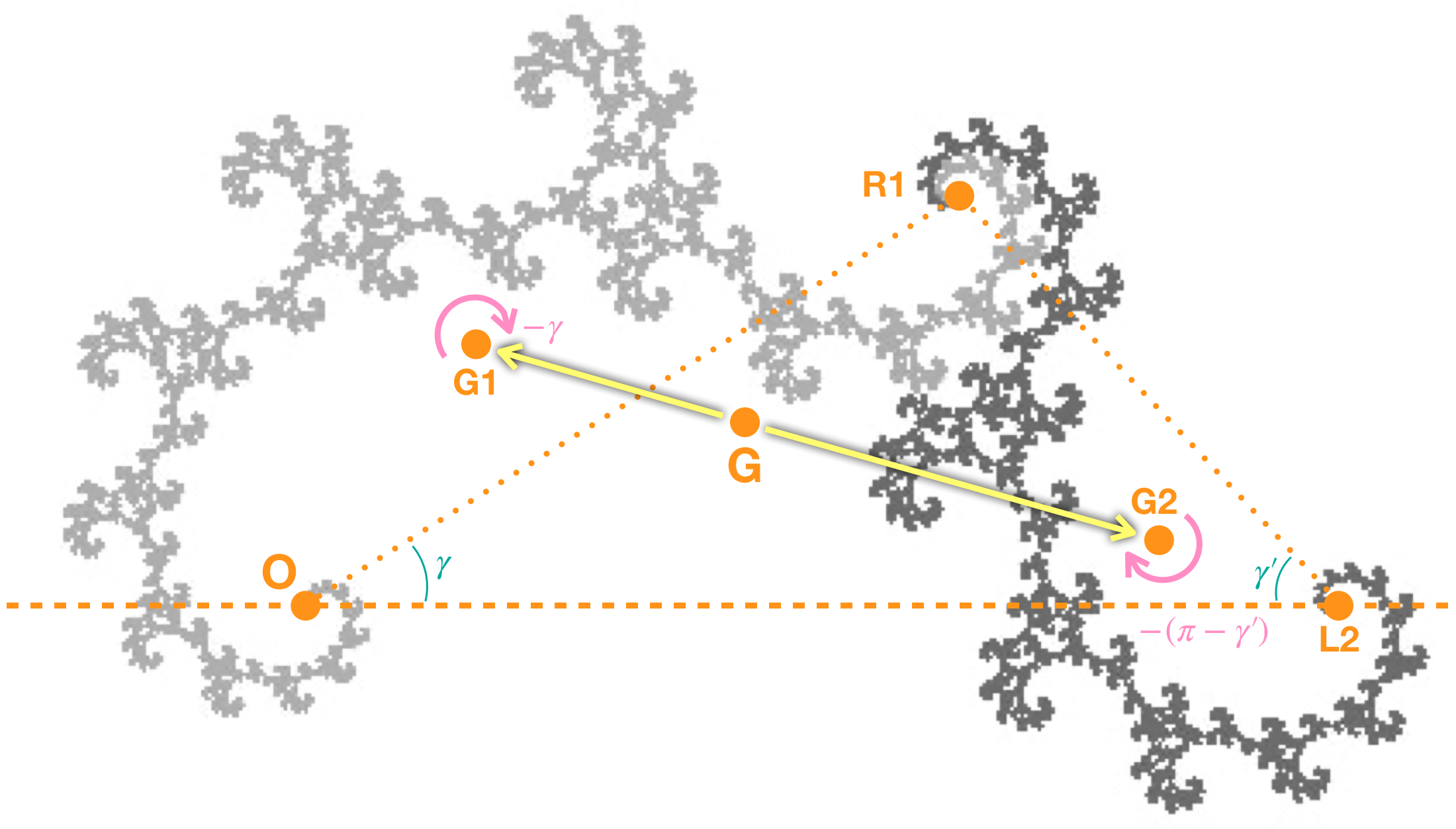}
\caption{\textbf{Golden dragon.} A golden dragon $\Omega^{(\text{Gd})}$ can be decomposed into two uneven smaller golden dragons $\Omega^{(\text{Gd}_1)} \cup \Omega^{(\text{Gd}_2)}$ as shown in two different shades of gray, which we call golden dragon $1$ (on the left) and golden dragon $2$ (on the right). Note that the rotation arrow indicates the chosen positive direction (clockwise), hence the angle values being negative here mean the smaller golden dragons are rotated in the negative direction (counter-clockwise).}
\label{fig03}
\end{figure*}

Golden dragon is an asymmetric fractal of similarity dimension $\mathscr{D}=\varphi =  \left(1+\sqrt{5}\right)/2 \approx 1.618$, which is the golden ratio. Choose a Cartesian coordinate system O$xy$ as in Fig. \ref{fig01}D, and let us normalize the distance between two heads $L\equiv 1$. A golden dragon {\color{black} $\Omega^{(\text{Gd})}$} can be decomposed into two uneven golden dragons {\color{black} that are self-similar to the original one}, which we call golden dragon $1$ and golden dragon $2$, {\color{black} i.e. $\Omega^{(\text{Gd}_1)}$ and $\Omega^{(\text{Gd}_2)}$}, as in Fig. \ref{fig03}. The right head of golden dragon $1$ is at position R$_1$ and the left head of golden dragon $2$ is at position L$_2$, which is also where the right head of the original golden dragon is. Let G, G$_1$, and G$_2$ be the centers of mass of the original golden dragon, golden dragon $1$, and golden dragon $2$. 

Define the distance between a pair of points A and A' to be $\overline{\text{AA'}}$. From the geometrical information as given in \cite{agnesscottGoldenDragon}, we can determine the distance between heads of golden dragon $1$ to be $L_1= \overline{\text{OR$_1$}} =\rho$ and the distance between heads of golden dragon $2$ to be $L_2=\overline{\text{R$_1$L$_2$}}=\rho^2$, where $\rho=(1/\varphi)^{(1/\varphi)}\approx0.74$. Use the law of cosines in trigonometry \cite{molokach2014law}, we can calculate the angle $\gamma$ and $\gamma'$ to be:
\begin{equation}
\gamma = \arccos \left( \frac{1+\rho^2 - \rho^4}{2\rho} \right) \approx 0.57 \ , \ \gamma' = \arccos \left( \frac{1+\rho^4 - \rho^2}{2\rho^2} \right) \approx 0.82  \ .
\end{equation}
Now we have known all parameters describing the geometric transformations from the original golden dragon to golden dragon $1$ and $2$. {\color{black} The spatial rescaling factor of $\Omega^{(\text{Gd}_1)}$ and $\Omega^{(\text{Gd}_2)}$ as compared to the original $\Omega^{(\text{Tw})}$ are not the same, i.e. $\eta_1 = L_1/L = \rho$ and $\eta_2 = L_2/L = \rho^2$. They also have different weights, i.e. $ W\left(\Omega^{(\text{Gd}_1)}\right) = \rho^\varphi$ and $ W\left(\Omega^{(\text{Gd}_2)}\right)= \rho^{2\varphi}$ as follows from Eq. \eqref{transf_chain_weight} and the weight normalization $W\left(\Omega^{(\text{Gd})}\right)=1$.}

We can get the smaller golden dragons from the original one by applying the following sequences of geometric transformation, in the same format with Eq. \eqref{transf_seq}:
\begin{equation}
    \Omega^{(\text{Gd}_1)}  = \mathcal{C}_{ [0,0 ];-\gamma;\rho}  \Omega^{(\text{Gd})} \ , \ \Omega^{(\text{Gd}_2)}  = \mathcal{C}_{ [1,0 ];-(\pi-\gamma');\rho^2}  \Omega^{(\text{Gd})} \ . 
    \label{get_gold}
\end{equation}

\ \ 

\subsection{Center of Mass}

Plug the transformation sequences as described in Eq. \eqref{get_gold} into Eq. \eqref{transf_chain_CoM} to get:
\begin{equation}
\begin{split}
    x_{\text{G}_1} = X\left( \Omega^{(\text{Gd}_1)}  \right) &= \rho \cos\gamma X\left( \Omega^{(\text{Gd})}  \right) - \rho \sin\gamma Y\left( \Omega^{(\text{Gd})}  \right) \ ,
    \\
    y_{\text{G}_1} = Y\left( \Omega^{(\text{Gd}_1)}  \right) &= \rho \sin\gamma X\left( \Omega^{(\text{Gd})}  \right) + \rho \cos\gamma Y\left( \Omega^{(\text{Gd})}  \right) \ ,
    \\
        x_{\text{G}_2} = X\left( \Omega^{(\text{Gd}_2)}  \right) &= -\rho^2 \cos\gamma' X\left( \Omega^{(\text{Gd})}  \right) - \rho^2 \sin\gamma' Y\left( \Omega^{(\text{Gd})}  \right) \ +1,
            \\
        y_{\text{G}_2} = Y\left( \Omega^{(\text{Gd}_2)}  \right) &=\rho^2 \sin\gamma' X\left( \Omega^{(\text{Gd})}  \right) - \rho^2 \cos\gamma' Y\left( \Omega^{(\text{Gd})}  \right) \ ,
\end{split}
\label{Gd_CoM_where}
\end{equation}
and also plug it into Eq. \eqref{transf_chain_weight}, then use these results into Eq. \eqref{additive_CoM} where we make the identifications $\Omega=\Omega^{(\text{Gd})}$, $\Omega_1=\Omega^{(\text{Gd}_1)}$, and $\Omega_2=\Omega^{(\text{Gd}_2)}$, we obtain a set of two consistency equations with two unknowns:
\begin{equation}
\begin{split}
    x_{\text{G}} = X\left( \Omega^{(\text{Gd})} \right) &= \rho^{2}\left(\cos{\gamma}-\rho^{2}\cos{\gamma'}\right) X\left( \Omega^{(\text{Gd})} \right) - \rho^{2}\left(\sin{\gamma}+\rho^{2}\sin{\gamma'}\right) Y\left( \Omega^{(\text{Gd})} \right) \ ,
    \\
        y_{\text{G}} = Y\left( \Omega^{(\text{Gd})} \right) &= \rho^{2}\left(\sin{\gamma}+\rho^{2}\sin{\gamma'}\right) X\left( \Omega^{(\text{Gd})} \right) + \rho^{2}\left(\cos{\gamma}-\rho^{2}\cos{\gamma'}\right)Y\left( \Omega^{(\text{Gd})} \right) \ .
        \end{split}
\label{Gd_ori_CoM_where}
\end{equation}
We can solve this analytically:
\begin{equation}
\begin{split}
    x_{\text{G}} &= \frac{\rho^{2\varphi} \left( \rho^{2\varphi} + \rho^2 \cos\gamma' \right)}{\rho^{4\varphi} + \rho^4 + 2 \rho^{2\varphi+2} \cos\gamma'}  \approx 0.39 \ ,
    \\
        y_{\text{G}} &= \frac{\rho^{2\varphi+2} \sin\gamma' }{\rho^{4\varphi} + \rho^4 + 2 \rho^{2\varphi+2} \cos\gamma'}  \approx 0.21 \ ,
\end{split}
\label{Gd_CoM}
\end{equation}
These are the first moments of the golden dragon fractal.

With the answers in Eq. \eqref{Gd_CoM}, we can plug these into Eq. \eqref{Gd_CoM_where} to find the locations of G$_1$ and G$_2$:
\begin{equation}
\begin{split}
    x_{\text{G}_1} &=  \frac{\rho^{2\varphi+1}\cos{\gamma} \left( \rho^{2\varphi} + \rho^2 \cos\gamma' \right)}{\rho^{4\varphi} + \rho^4 + 2 \rho^{2\varphi+2} \cos\gamma'}x_{\text{G}} - \frac{\rho^{2\varphi+3}\sin{\gamma} \sin\gamma' }{\rho^{4\varphi} + \rho^4 + 2 \rho^{2\varphi+2} \cos\gamma'}y_{\text{G}} \approx 0.16 \ ,
    \\
    y_{\text{G}_1} &=  \frac{\rho^{2\varphi+1}\sin{\gamma} \left( \rho^{2\varphi} + \rho^2 \cos\gamma' \right)}{\rho^{4\varphi} + \rho^4 + 2 \rho^{2\varphi+2} \cos\gamma'}x_{\text{G}} + \frac{\rho^{2\varphi+3}\cos{\gamma} \sin\gamma' }{\rho^{4\varphi} + \rho^4 + 2 \rho^{2\varphi+2} \cos\gamma'}y_{\text{G}} \approx 0.29 \ ,
    \\
        x_{\text{G}_2} &=  -\frac{\rho^{2\varphi+2}\cos{\gamma'} \left( \rho^{2\varphi} + \rho^2 \cos\gamma' \right)}{\rho^{4\varphi} + \rho^4 + 2 \rho^{2\varphi+2} \cos\gamma'}x_{\text{G}} - \frac{\rho^{2\varphi+4} \sin^{2}\gamma' }{\rho^{4\varphi} + \rho^4 + 2 \rho^{2\varphi+2} \cos\gamma'}y_{\text{G}}+1 \approx 0.77 \ ,
            \\
        y_{\text{G}_2} &=  \frac{\rho^{2\varphi+2}\sin{\gamma'} \left( \rho^{2\varphi} + \rho^2 \cos\gamma' \right)}{\rho^{4\varphi} + \rho^4 + 2 \rho^{2\varphi+2} \cos\gamma'}x_{\text{G}} - \frac{\rho^{2\varphi+4} \sin{\gamma'}\cos{\gamma'} }{\rho^{4\varphi} + \rho^4 + 2 \rho^{2\varphi+2} \cos\gamma'}y_{\text{G}} \approx 0.080 \ .
\end{split}
\label{Gd1_Gd2_CoM}
\end{equation}

\ \ 

\subsection{Moments of Inertia}

We use the transformation sequences Eq. \eqref{get_gold} in Eq. \eqref{transf_chain_weight} and Eq. \eqref{transf_chain_MoI}, then plug these findings into Eq. \eqref{useful_rel} where we make the identifications $\Omega=\Omega^{(\text{Gd})}$, $\Omega_1=\Omega^{(\text{Gd}_1)}$, and $\Omega_2=\Omega^{(\text{Gd}_2)}$. For the values of $\left[x_{\text{G}},y_{\text{G}} \right]$, $x_{\text{G}}$, $\left[x_{\text{G}_1},y_{\text{G}_1} \right]$, and $\left[x_{\text{G}_2},y_{\text{G}_2} \right]$, we can utilize our answers in Eq. \eqref{Gd_CoM} and \eqref{Gd1_Gd2_CoM}. Finally, after some exhausting algebraic manipulations we have a set of three consistency equations with three unknowns:
\begin{equation}
\begin{split}
\tilde{I}_{xx} \left( \Omega^{(\text{Gd})} \right) &= (1-a) \tilde{I}_{xx} \left( \Omega^{(\text{Gd})} \right) + b\tilde{I}_{yy} \left( \Omega^{(\text{Gd})} \right)  -2c \tilde{I}_{xy} \left( \Omega^{(\text{Gd})} \right) + X_1 \ ,
\\
\tilde{I}_{yy} \left( \Omega^{(\text{Gd})} \right) &= b \tilde{I}_{xx} \left( \Omega^{(\text{Gd})} \right) + (1-a)\tilde{I}_{yy} \left( \Omega^{(\text{Gd})} \right) + 2c\tilde{I}_{xy} \left( \Omega^{(\text{Gd})} \right) + X_2 \ ,
\\
\tilde{I}_{xy} \left( \Omega^{(\text{Gd})} \right) &= c \tilde{I}_{xx} \left( \Omega^{(\text{Gd})} \right) -c\tilde{I}_{yy} \left( \Omega^{(\text{Gd})} \right) + (1-a-b) \tilde{I}_{xy} \left( \Omega^{(\text{Gd})} \right) + X_3 \ ,
\end{split}
\label{Gd_consistency_eqs}
\end{equation}
in which we have used the coefficients $a$, $b$, and $c$:
\begin{equation}
    a=\rho^{2\varphi}+2\rho^{\varphi+2}\cos \gamma '-\rho^4\cos^2\gamma ' \ , \ b=\rho^4\sin^2\gamma ' \ , \ 
    c=\rho^2\sin \gamma '(\rho^\varphi -\rho^2\cos \gamma ') \ .
    \label{final_phantasy}
\end{equation}
{\color{black} We derive these equations in Appendix \ref{Gd_consistency}. Then, let us define:}
\begin{equation}
    K = \dfrac{\rho^{2\varphi}}{\rho^4+\rho^{4\varphi}+2\rho^{2\varphi+2}\cos \gamma '} \ ,
\end{equation}
then the terms $X_1$, $X_2$, and $X_3$ are given by:
\begin{equation}
\begin{split}
    X_1=K^2\rho^{3\varphi+4}\sin^2\gamma ' \ , \ X_2=K^2(\rho^4+\rho^{2\varphi+2}\cos \gamma ')^2(1+\rho^{\varphi}) \ ,
    \\
    \ X_3=K^2\rho^{\varphi+2} \sin \gamma '(\rho^4+\rho^{2\varphi+2}\cos \gamma ') \ .
    \end{split}
    \end{equation}
We can further proceed without much difficulty, given that these equations in Eq. \eqref{final_phantasy} are all linear, to obtain analytical solutions:
\begin{equation}
\begin{split}
\tilde{I}_{xx} \left( \Omega^{(\text{Gd})} \right) &= \dfrac{(a^2+ab+2c^2)X_1+(b^2+ab+2c^2)X_2-2c(a-b)X_3}{(a-b)[(a+b)^2+4c^2]} \approx 0.040 \ ,
\\
\tilde{I}_{yy} \left( \Omega^{(\text{Gd})} \right) &= \dfrac{(a^2+ab+2c^2)X_2+(b^2+ab+2c^2)X_1+2c(a-b)X_3}{(a-b)[(a+b)^2+4c^2]} \approx 0.14 \ ,
\\
\tilde{I}_{xy} \left( \Omega^{(\text{Gd})} \right) &= \dfrac{c(X_1-X_2)+(a+b)X_3}{(a+b)^2+4c^2} \approx 0.023 \ ,
\end{split}
\end{equation}
These are the second moments of the golden dragon fractal.

\ \ 

\section{Comparison with Numerical Evaluations}

We can do similar calculations for terdragon $\Omega^{(\text{Te})}$ (see Fig. \ref{fig01}B) and Heighway dragon $\Omega^{(\text{He})}$ (see Fig. \ref{fig01}C). Note that we have to decompose Heighway dragon into three smaller Heighway dragons instead of two. The centers of mass are at positions:
\begin{equation}
\left[ X\left( \Omega^{(\text{Te})}\right), Y\left( \Omega^{(\text{Te})}\right)\right] = \left[ 0,0 \right] \ , \ \left[ X\left( \Omega^{(\text{He})}\right), Y\left( \Omega^{(\text{He})}\right)\right] = \left[ \frac25,\frac15 \right] \ .
\end{equation}
The moments of inertia are found to be:
\begin{equation}
\tilde{\mathbb{I}}\left( \Omega^{(\text{Te})}\right) = \begin{bmatrix}
\frac1{42} & \frac1{28\sqrt{3}}  \\
\frac1{28\sqrt{3}} & \frac5{48} 
\end{bmatrix} \ , \ \tilde{\mathbb{I}}\left( \Omega^{(\text{He})}\right) = \begin{bmatrix}
\frac3{50} & \frac3{100} \\
\frac3{100} & \frac7{50} 
\end{bmatrix} \ .
\end{equation}
These are the analytical answers for the first and second moments of terdragon and Heighway dragon.

We compare our theoretical findings with the estimations obtained from running numerical calculations on MatLab \cite{MATLAB}, which have been analyzed on pixelated images of these dragons. In essence, for an image of a fractal dragon (typically $1200 \times 2000$ pixels), we consider each pixel containing a portion of the fractal as a constituent positioned at the center of that pixel, associated with a weight equal to $W\equiv 1$ divided by the total number of pixels the fractal is made of. The digital distance between two pixels correspond to two heads is normalized to $L\equiv 1$. We then numerically calculate the center of mass and the moments of inertial for this effective description of the fractal dragon, using Eq. \eqref{CoM_eq} and Eq. \eqref{MoI_eq}. The MatLab codes and images of fractal dragons for estimating the first and second moments will be provided by the corresponding author upon requests.

We see good agreements between theory and numerical evaluations, as shown in Table \ref{table:1}.

\begin{table}[!ht]
\centering
\begin{tabular}{ | m{8em} | m{2.4cm}| m{2.4cm} | m{2.4cm} | m{2.4cm} | m{2.4cm} | }
  \hline
  \makecell{Fractal} & \makecell{$x_\text{G}$} & \makecell{$y_\text{G}$} & \makecell{$\tilde{I}_{xx}$} & \makecell{$\tilde{I}_{yy}$} & \makecell{$\tilde{I}_{xy}$} \\ 
  \hline
  \makecell{Twindragon \\ (Fig. \ref{fig01}A)} & \makecell{0 (theo) \\ 0 (num)} & \makecell{0 (theo) \\ 0 (num)} & \makecell{0.10 (theo) \\ 0.11 (num)} & \makecell{0.15 (theo) \\ 0.16 (num)} & \makecell{0.050 (theo) \\ 0.05 (num)} \\ 
  \hline
  \makecell{Terdragon \\ (Fig. \ref{fig01}B)} & \makecell{0 (theo) \\ 0 (num)} & \makecell{0 (theo) \\ 0 (num)} & \makecell{0.023 (theo) \\ 0.024 (num)} & \makecell{0.060 (theo) \\ 0.062 (num)} & \makecell{0.021 (theo) \\ 0.021 (num)} \\ 
  \hline
   \makecell{Heighway dragon \\ (Fig. \ref{fig01}C)} & \makecell{0.40 (theo) \\ 0.40 (num)} & \makecell{0.20 (theo) \\ 0.19 (num)} & \makecell{0.060 (theo) \\ 0.063 (num)} & \makecell{0.14 (theo) \\ 0.15 (num)} & \makecell{0.030 (theo) \\ 0.031 (num)} \\  
  \hline
   \makecell{Golden dragon \\ (Fig. \ref{fig01}D)} & \makecell{0.39 (theo) \\ 0.38 (num)} & \makecell{0.21 (theo) \\ 0.21 (num)} & \makecell{0.040 (theo) \\ 0.041 (num)} & \makecell{0.14 (theo) \\ 0.14 (num)} & \makecell{0.023 (theo) \\ 0.023 (num)} \\  
  \hline
\end{tabular}
\caption{We apply our method of calculating the first and second moments for complex dragon fractals. Comparison between pen-and-paper theoretical results (theo) with computer numerical findings (num) show only small deviations.}
\label{table:1}
\end{table}

\section{Discussion}

We have demonstrated how the mechanical properties of fractals made of fractals, such as the dragons \cite{agnesscottTwindragon,agnesscottTerdragon,agnesscottHeighwayDragon,agnesscottGoldenDragon}, can be calculated by our proposed method which involves reconstructing the original shape through geometrically transformed versions of itself. {\color{black} We can use similar procedure to study other fractals, e.g. the Rauzy fractals \cite{ito1991rauzy} and the recently proposed Kochawave tiles \cite{sigrist2022kochawave}.} While we currently have derived only the first two moments of these dragon fractals, which correspond to the center of mass and the moments of inertia, it is feasible to apply the same analysis to higher moments. {\color{black} However, a potential limitation of our approach is its inability to easily apply to properties that do not transform as tensors under rotation -- further investigation is needed in this regard.} Another direction worth heading is  toward the transition from lower to higher dimensions, e.g. moving from fractals embedded in two-dimensional space to those existing in three-dimensional space. Examples include Sierpinski's Platonic fractals or three-dimensional dragons with their bodies writhing out of the plane. {\color{black} It is also curious to study possible external disturbances and geometric uncertainties in the designs of these fractal structures that could lead to chaotic dynamic \cite{pal2022adaptive}.}

In many {\color{black} common} mathematical playgrounds {\color{black} for physicists}, advanced topics in number theory \cite{aref2007point,do2022equal} and classical geometry \cite{misner1957classical} are often seen as abstract and distanced from immediate engineering applications. However, {\color{black} at} the intersection between these disciplines, fractals offer an opportunity for exploration beyond mere recreational challenges, propelled by current technological advancements \cite{phan2020bacterial}. While our focus here is on investigating how the geometrical information of fractals can help determining their mechanical properties, we intend is to extend this exploration to other physical properties e.g. thermodynamics, hydrodynamics, electronics, and electromagnetics. These properties have been of interests in various engineering sectors \cite{levy2005fractals}. Also, the concept of fractals goes beyond spatial dynamics, and finds its manifestation in temporal dynamics as well. It has proven to be useful in handling time series of complex statistics, by modeling the signal generators or control systems as obeying fractal time derivative equations \cite{li2010fractal}.

In summary, fractal physics has provided a compelling avenue for further exploration. Within recent technological advances, the capability to design and create fractals has already caught up, opening up new possibilities for many engineering applications.

\section{Acknowledgement}

We thank the xPhO club for their support to share this finding to a wider audience, and many useful comments to make this paper more approachable for general readers.

\ \ 

{\color{black}
\appendix

\section{On the Scale Transformation of the Second-Moment \label{scale_transf}}

Consider the change in the spatial scale, with rescaling factor $\eta$. Physically, the second-moment represents moment of inertial, which has the dimension of mass (weight) times area (length-squared). As follow from Eq. \eqref{transf_weight}, the weight is rescaled by $\eta^{\mathscr{D}}$ from the definition of similar dimension. Multiplying that with $\eta^2$ for the rescaling factor of length-squared, we get the total factor of $\eta^{\mathscr{D}+2}$ as shown in Eq. \eqref{transf_scale_MoI}.

\section{On the Consistency Relations}

\subsection{Twin Dragon \label{Tw_consistency}}

Let us start with the $xx$-component of the twin-dragon moment of inertia evaluated at its center of mass, which can be represented in two ways -- as itself $\tilde{I}_{xx} \left( \Omega^{(\text{Tw})} \right)$, or as the summation of two contributions from smaller twin-dragons i.e. $\Omega^{(\text{Tw}_1)}$ and $\Omega^{(\text{Tw}_1)}$. The twin dragons can be transformed into by the original twin dragon via Eq. \eqref{Tw1_and_Tw2_transf}, which we can then apply Eq. \eqref{transf_chain_MoI} to get their contribution to the $xx$-component:
\begin{equation}
\begin{split}
\tilde{I}_{xx}\left( \Omega^{(\text{Tw}_1)} \right) \ \ \overset{\text{use Eq. \eqref{Tw1_and_Tw2_transf}}}{=} & \ \  \tilde{I}_{xx}\left( \mathcal{C}_{ [ -2^{-2},2^{-2}];-2^{-2}\pi;2^{-1/2}}  \Omega^{(\text{Tw})} \right) 
\\
 \overset{\text{use Eq. \eqref{transf_chain_MoI}}}{=} & \ \ 2^{-3} \tilde{I}_{xx}\left(\Omega^{(\text{Tw})}\right) + 2^{-3} \tilde{I}_{yy}\left(\Omega^{(\text{Tw})}\right) + 2^{-2} \tilde{I}_{xy}\left(\Omega ^{(\text{Tw})}\right) \ \ .
\\
\tilde{I}_{xx}\left( \Omega^{(\text{Tw}_2)} \right) \ \ \overset{\text{use Eq. \eqref{Tw1_and_Tw2_transf}}}{=} & \ \  \tilde{I}_{xx}\left( \mathcal{C}_{ [ 2^{-2},-2^{-2}];-2^{-2}\pi;2^{-1/2}}  \Omega^{(\text{Tw})} \right) 
\\
 \overset{\text{use Eq. \eqref{transf_chain_MoI}}}{=} & \ \ 2^{-3} \tilde{I}_{xx}\left(\Omega^{(\text{Tw})}\right) + 2^{-3} \tilde{I}_{yy}\left(\Omega^{(\text{Tw})}\right) + 2^{-2} \tilde{I}_{xy}\left(\Omega ^{(\text{Tw})}\right) \ \ .
\end{split}
\label{xx_2part}
\end{equation}
Note that we also use $\mathscr{D} = 2$ for the similar dimensionality of a twin dragon. The smaller twin dragons have equal weights, i.e. $ W\left(\Omega^{(\text{Tw}_1)}\right) = W\left(\Omega^{(\text{Tw}_2)}\right)= 1/2$, which allows us to utilize Eq. \eqref{useful_rel} to arrive at the consistency equation for $xx$-component:
\begin{equation}
    \tilde{I}_{xx} \left( \Omega^{(\text{Tw})} \right) = \tilde{I}_{xx} \left( \Omega^{(\text{Tw}_1)} \right) + \tilde{I}_{xx} \left( \Omega^{(\text{Tw}_2)} \right) + W\left( \Omega^{(\text{Tw}_1)} \right)  y_\text{G$_1$}^2 + W\left( \Omega^{(\text{Tw}_2)} \right) y_\text{G$_2$}^2
\label{xx_move}
\end{equation}
Here we have used $\left[ x_{\text{G}},y_{\text{G}}\right]=\left[ 0, 0\right]$. Since $\left[ x_{\text{G}_1},y_{\text{G}_1}\right]=\left[ - 2^{-2}, 2^{-2}\right]$ and  $\left[ x_{\text{G}_2},y_{\text{G}_2}\right]=\left[ 2^{-2}, - 2^{-2}\right]$, we can then plug Eq. \eqref{xx_2part} into Eq. \eqref{xx_move} to obtain:
\begin{equation}
    \tilde{I}_{xx} \left( \Omega^{(\text{Tw})} \right) = 2^{-2} \tilde{I}_{xx} \left( \Omega^{(\text{Tw})} \right) + 2^{-2} \tilde{I}_{yy} \left( \Omega^{(\text{Tw})} \right) - 2^{-1} \tilde{I}_{xy} \left( \Omega^{(\text{Tw})} \right) + 2^{-4} \ \ .
\label{Tw_xx_consistency_eqs}
\end{equation}
Repeat the steps above for the $yy$- and $xy$-components, we can get three consistency equations as in Eq. \eqref{Tw_consistency_eqs}.

\subsection{Golden Dragon \label{Gd_consistency}}

The derivation of Eq. \eqref{Gd_consistency_eqs} is very close to Appendix \ref{Tw_consistency} with just a few slight differences. We can also start with the $xx$-components of the moment of inertial $\tilde{I}_{xx} \left( \Omega^{(\text{Gd})} \right)$ as in Eq. \eqref{xx_2part}, but the transformation from the original dragon $\Omega^{(\text{Gd})}$ to smaller dragons $\Omega^{(\text{Gd}_1)}$ and $\Omega^{(\text{Gd}_2)}$ are Eq. \eqref{get_gold} instead of Eq. \eqref{Tw1_and_Tw2_transf}. These smaller golden dragons are not of equal size and weights, i.e. $W\left(\Omega^{(\text{Gd}_1)}\right) = \rho^\varphi$ and $ W\left(\Omega^{(\text{Gd}_2)}\right)= \rho^{2\varphi}$, and the positions for the centers of masses are more elaborated, i.e. Eq. \eqref{Gd_CoM_where} and Eq. \eqref{Gd_ori_CoM_where}. With these notices, we can follow Eq. \eqref{xx_move} and Eq. \eqref{Tw_xx_consistency_eqs} to obtain the consistency equation in $xx$-components, then do the same for the other two components i.e. $yy$- and $xy$-, thus finally arrive at Eq. \eqref{Gd_consistency_eqs}.

}

\bibliography{main}

\end{document}